\documentclass[aps, prl, twocolumn, showkeys, floatfix]{revtex4-2}

\usepackage{chemformula} % for chemical symbols and formulae
\usepackage{siunitx} % for si units
\usepackage[inline,shortlabels]{enumitem} % inline numbered lists
\usepackage{hyperref} %hyperref can cause crash if citation is split over two pages. This is remedied by the [draft] option, which can be removed after rearranging the text a bit.
\usepackage{graphicx} % including graphics
\usepackage{tabularx}
\usepackage{threeparttablex} % Tables with notes and citations
\usepackage[section]{placeins}
\usepackage{multirow}
\usepackage{fancyhdr}
\usepackage{dblfloatfix}
\usepackage{float}
\usepackage{caption}
\usepackage{subcaption}

\hypersetup{hidelinks}

\makeatletter
\renewcommand{\fnum@figure}{Fig. \thefigure}
\makeatother

\usepackage[justification=raggedright, labelsep=period]{caption}
\begin{document}

    \title{High-Density Polyethylene Degradation into Low Molecular Weight Gases at 1823 K: An Atomistic Simulation}
    \author{Leyla Ramin}
    \affiliation{Centre for Sustainable Materials Research and Technology (SMaRT), School of Materials Science and Engineering, University of New South Wales, Sydney NSW 2052, Australia.}
    \author{M. Hussein N. Assadi}
    \affiliation{Centre for Sustainable Materials Research and Technology (SMaRT), School of Materials Science and Engineering, The University of New South Wales, Sydney, 2052, Australia}
    \affiliation{Graduate School of Engineering Science, Osaka University, Toyonaka, Osaka 560-8531, Japan.\\
    \url{assadi@aquarius.mp.es.osaka-u.ac.jp}}
    \author{Veena Sahajwalla}
    \affiliation{Centre for Sustainable Materials Research and Technology (SMaRT), School of Materials Science and Engineering, The University of New South Wales, Sydney, 2052, Australia.}
    \date{2014}
    
    \begin{abstract}
    Using molecular dynamics simulation, we present a comprehensive study of the volatile thermal degradation of high-density polyethylene (HDPE) across a temperature range of 300 K to 1823 K. We find that degradation at temperatures higher than $\sim 1373$ K generates significant quantities of reducing gases such as \ch{CH_n} and hydrogen molecules which are beneficial to the steelmaking industry. Our results provide a new understanding of HDPE's phase transformation from solid to gas that occurs during superheating at steelmaking's electric arc furnace environment offering a new method for eliminating end-of-life HDPE from landfill.
    \end{abstract}

\maketitle

       \section{INTRODUCTION}%\FloatBarrier
The massive increase in the volume of end-of-life polymeric products is driving intense research efforts towards more innovative recycling techniques [1--3]. For instance, one emerging opportunity is to use waste polymers to generate fuel sources and reducing agents for steelmaking industry through high temperature degradation process [4,5]. In addition to reducing the volume of the landfilled waste polymers, this approach has the economic benefit of partially substituting expensive conventional fossil fuels. As an advancement in this field, we have recently achieved enhanced volume and improved stability for slag when we used a blend of high–density polyethylene (HDPE) and natural coke in a steelmaking electric arc furnace [5]. The improved slag quality was attributed to the kinetic advantage of using HDPE as it produces higher hydrogen output during high temperature degradation which enhanced slag foaming.\par
Although, thermal degradation of HDPE has been studied both experimentally [6-8] and theoretically [9], these investigations were limited to temperatures below ~ 973 K. Therefore, the atomistic description of the high temperature ($\sim 1800$ K) degradation of HDPE during rapid heating has not been fully investigated yet. The main reason for this is the challenge posed by the prohibitively high temperatures and large scale settings [10] that inhibits probing of all thermo-chemical reactions individually [11]. Furthermore, theoretical mechanistic [12] and empirical [13] models that have previously been developed to predict the final products of gasification and pyrolysis of polymers have not considered temperatures higher than $\sim 1273$ K [14]. Therefore atomistic molecular dynamics (MD) simulation appears to be an insightful tool for studying high temperature degradation of the polymeric structures and the formation of the final products. In this work, we use comprehensive MD simulation to study thermal degradation of HDPE at temperatures as high as 1823 K.\par

       \section{Simulation Settings}%\FloatBarrier
We used the molecular dynamics package Large-scale Atomic/Molecular Massively Parallel Simulator (LAMMPS) [15] to simulate the high temperature degradation of HDPE. A cubic cell containing a single chain HDPE with 80 C atoms and 162 H atoms was used as the initial geometry. This approach allows studying HDPE's decomposition in reductive environment similar to the one of the electric arc furnace (EAF). Using a relatively a large molecular chain also allows detailed evaluation of the chemical reactions associated with the molecular transformation process. We then applied periodic boundary condition (PBC) to the system in X, Y and Z directions. PBC allows the molecules near the either boundaries to interact with the molecules of the opposite sides and thus facilitates realistic simulation without being computationally too demanding. All interactions between the C and H atoms including Coulomb forces were governed by a Reax potential [16,17]. Reax Potential is a general bond order based force field that provides precise descriptions of the bond breaking and bond forming process in hydrocarbons. Reax potential accounts for nonbonding interactions such as van der Waals and Coulomb forces by considering interactions between every pair of atoms, irrespective of connectivity. The initial random velocities for individual atoms were assigned to give a Maxwell--Boltzmann distribution corresponding to the target temperature. The equations of motion were integrated using the velocity Verlet algorithm with a time step of 0.03 fs.\par
We used the following algorithms to reach equilibrium state at any given temperature: (1) NVE integration where number of moles (N), volume (V) and energy (E) are kept constant [18]. Within NVE ensemble the temperature is raised by adding non–translational kinetic energy i.e. heat to the equilibrated configuration in a manner that conserves atoms aggregate momentum; (2) Langevin thermostat to recalculate the position of the atoms at each time step [19]; (3) Berendsen barostat to stabilize the pressure fluctuations in the system with the target pressure of 1 atm [20]. To see if the simulation had reached equilibrium, we monitored the energy, temperature and pressure fluctuations at regular intervals during the simulation. By $t = 1.5$ ns the system reached stable condition where the fluctuations of energy, temperature and pressure became minimal.\par
       
       \section{RESULTS AND DISCUSSION}%\FloatBarrier
       \subsection{Initial Stages of Degradation}

Using the equilibrated configuration, we applied heating rate of 1.5 kcal$/$mol.fs to increase the temperature from 300 K to 1823 K to investigate the thermal transformation of HDPE as a function of temperature. The applied temperature range extends across ambient to the temperature of molten iron in an EAF. First we present the calculated density of HDPE chain molecule and the resultant products during heating as plotted in Fig. 1(a). The density at 300 K for HDPE chain in this simulation is $\sim 640$ Kg$/$m$^3$ which is in agreement with the measured values of bulk HDPE density in pellet form [21]. Then as temperature increased to 673 K (Stage I) the density decreased monotonically. However for temperatures between 437 K and 562 K (Stage II) the rate of density change almost stagnated. The change in the density slopes between Stage I and Stage II is comparable to the solid/liquid HDPE density ratio which indicates solid to liquid phase transition. Experimentally, the melting point of HDPE was reported to be in the range of 493 -- 553 K depending on the HDPE's manufacturing process. Therefore the molecular dynamics approach, as employed here, reproduces the experimental melting point [22]. Furthermore, Stage II occurred over a very small temperature range as the time spent in liquid phase is minimal due to the rapid heating. In stage III, the density started to decrease with a lower rate when compared with Stages I and II indicating the start of phase transition from liquid to gas. Once the temperatures passed the $\sim 1373$ K mark, the density diminished greatly and then remained constant at $\sim 2.4$ Kg$/$m$^3$ suggesting the products were completely in gaseous phase.\par

          \begin{figure}[bth!]
            \centering
           \begin{subfigure}[t]{0.42\textwidth}
           \centering
           \includegraphics[width=\textwidth]{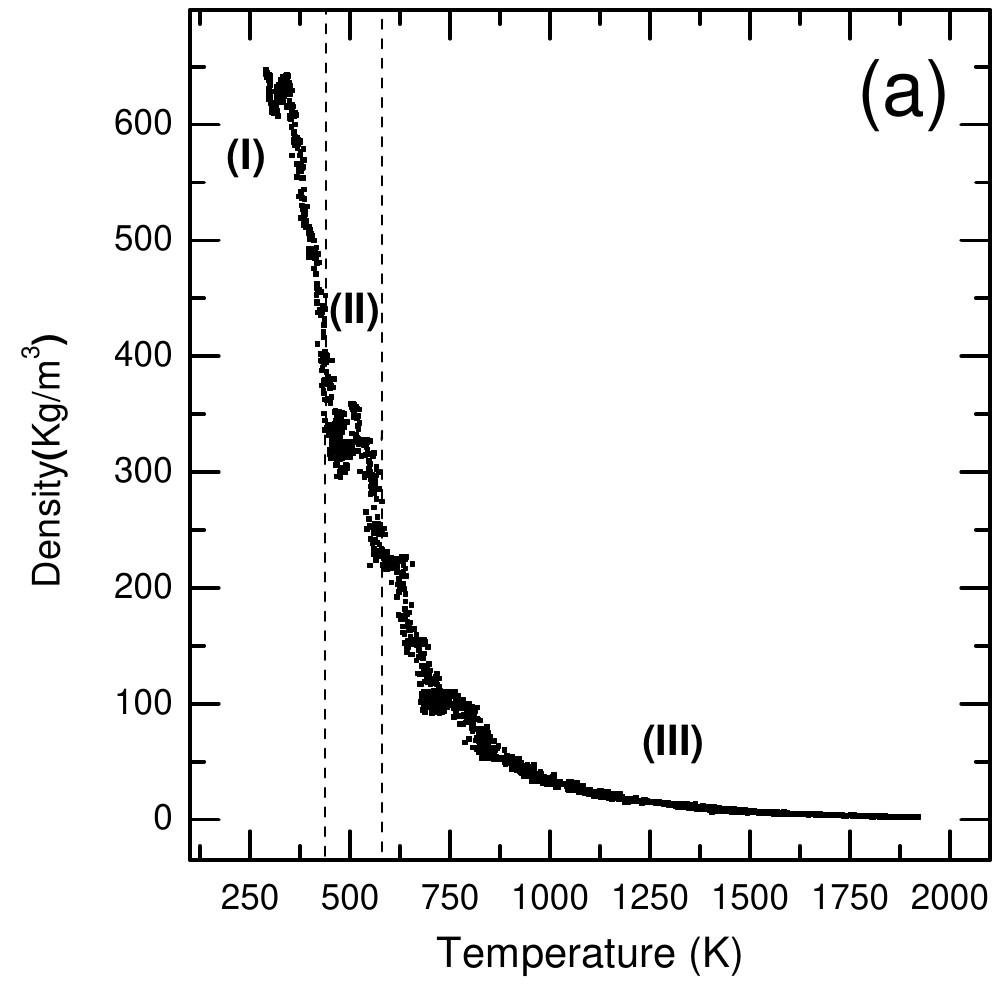} 
          \end{subfigure}
          
          \begin{subfigure}[t]{0.42\textwidth}
           \centering
           \includegraphics[width=\textwidth]{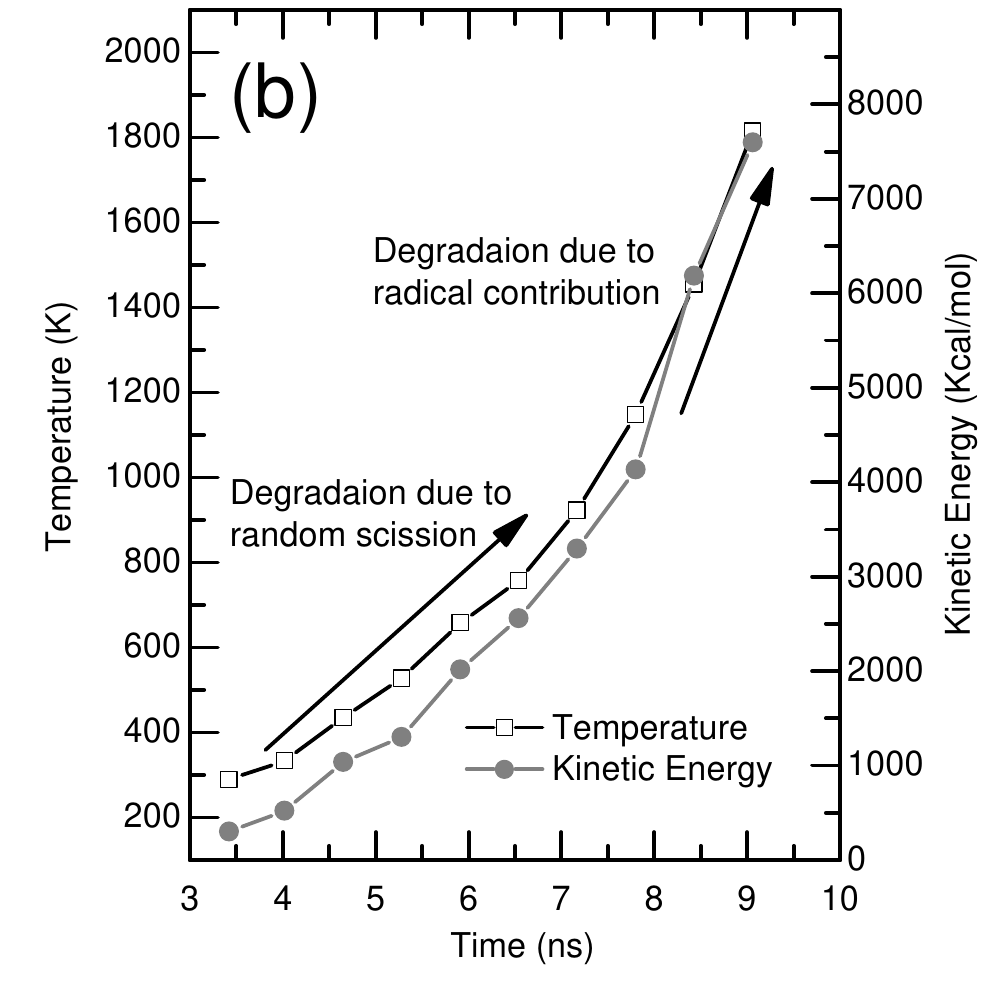} 
          \end{subfigure}
           \caption{\label{figure:1}(a) The calculated density as function of temperature. (b) The evolution of kinetic energy of HDPE chain and temperature as a function of simulation time.}
         \end{figure}

With increasing temperature, we counted the resultant smaller C and H chains by examining the bond connectivity at various temperatures to monitor the structural changes as a function of temperature. The results are presented in Table 1. We found that the HDPE chain maintained its structural integrity at temperatures up to $\sim 330$ K. For higher temperature the C--C and C--H bonds started to vibrate with higher amplitudes although no bond broke for a while. At 703 K, the main chain began to break down into few fresh HDPE chain molecules through scissions. The prediction of the HDPE chain breakage is consistent with earlier thermogravimetric analysis reporting that the degradation of HDPE begins to take place in the 703 K -- 743 K temperature range [23]. Furthermore, isothermal and dynamic decomposition measurement of HDPE also detected that the chemical transformation of HDPE starts at $T = \sim 723$ K [24].\par
For temperatures above 766 K, the HDPE monomers further fragmented into few hydrocarbon molecules, all containing four or more C atoms. As seen in Table 1, at this temperature range, the dominant outcome of the degradation process was generally heavier molecules; for instance, at $T = 849$ K, there were only two lighter \ch{C2H_n} molecules while there were four bigger hydrocarbon chains. This trend is on agreement of experimental pyrolysis of polyethylene, in which at 873 K, HDPE yielded negligible amount of gaseous products [8]. When temperature increased to 1396 K, on the other hand, there was no hydrocarbon molecule containing more than three C atoms whereby the absence of larger molecules indicated the complete transformation into the gaseous phase which is in agreement with the density calculations of Fig. 1(a). One noteworthy point was that all resultant molecules were linear, indicating the absence of toxic polycyclic aromatic hydrocarbons in the final products.
                       
       \subsection{Diffusion Process}

           \begin{table}[bth!]
\caption{Number of molecules generated from HDPE molecular transformation obtained by MD simulation.}
\begin{center}
\begin{threeparttable}
\begin{tabular}{l c c c c c}
  \hline\hline
Temperature (K) & \ch{C_{80-4}H_n} & \ch{C3H_n} & \ch{C2H_n} & \ch{CH_n} & \ch{H2} \\
  \hline
1823 & --- & 7 & 20 & 19 & 9 \\
1735 & --- & 5 & 22 & 21 & 8 \\
1649 & --- & 8 & 21 & 14 & 5 \\
1581 & --- & 8 & 21 & 14 & 4 \\
1567 & --- & 8 & 22 & 12 & 2 \\
1537 & --- & 8 & 21 & 14 & 2 \\
1455 & --- & 8 & 22 & 12 & 2 \\
1396 & --- & 12 & 19 & 6 & 1 \\
1333 & 1 & 11 & 18 & 4 & --- \\
1222 & 4 & 7 & 16 & 3 & --- \\
1130	 & 5 & 6 & 11 & 3 & --- \\
1057 & 5 & 4 & 8 & 2 & --- \\
988 & 5 & 3 & 8 & 2 & --- \\
930 & 4 & 1 & 5 & --- & --- \\
899 & 4 & --- & 3 & --- & --- \\
849 & 4 & --- & 2 & --- & --- \\
805 & 3 & --- & 2 & --- & --- \\
766 & 3 & --- & 1 & --- & --- \\
723 & 3 & --- & --- & --- & --- \\
706 & 2 & --- & --- & --- & --- \\
704 & 1 & --- & --- & --- & --- \\
700 & 1 & --- & --- & --- & --- \\
  \hline\hline
\end{tabular}
\end{threeparttable}
\end{center}
\label{table:1}
\end{table}
         
\begin{figure}[bth!]
            \centering
            \includegraphics[width=1\linewidth]{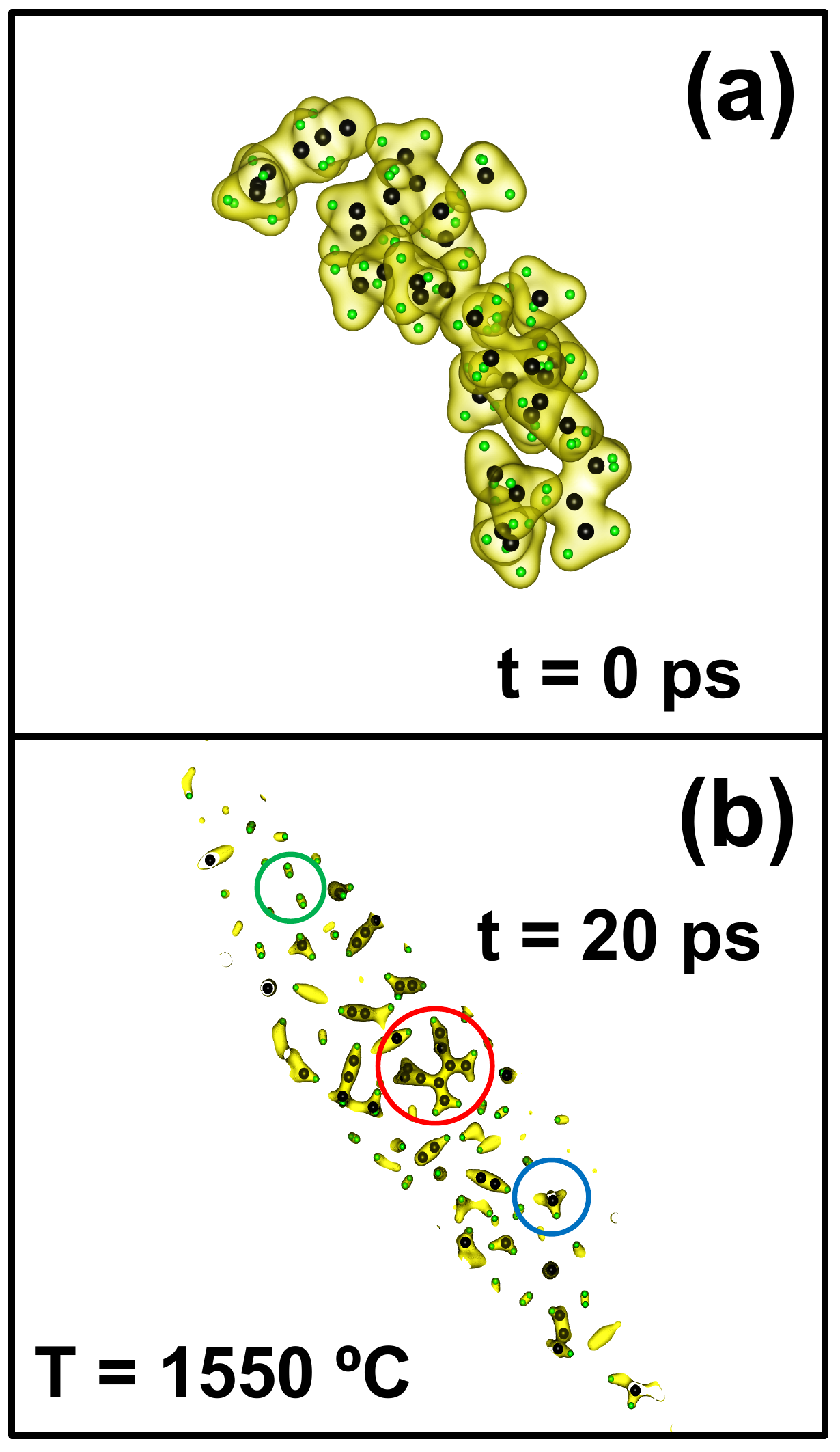}
            \caption{\label{figure:2}(a) The schematic representation of intact high-density polyethylene chain molecule at room temperature. Larger black spheres represent C atoms while smaller gray spheres represent H atoms. The shade around the atoms represents charge densities (b) The interacting light molecules generated at $T = 1823$ K. Sample \ch{H2} and \ch{CH4} molecules are marked by circles.}
\end{figure}

Fig. 1(b) shows the average kinetic energy as a function of temperature. For temperatures below $\sim 1373$ K, the kinetic energy of the system was composed of the kinetic energy of the randomly scissioned radicals. Since, all the C--C bonds in polyethylene have the same strength (except those in the terminal positions), HDPE chain initially decomposed by a random-chain scission mechanism. As the random scission process continued, the freshly produced radicals further interacted with the remaining HDPE chain molecule and sped up the gasification process. As temperature rose, degradation by radicals became dominant over the scission mechanism. Since degradation by radicals requires higher kinetic energy, the value of kinetic energy of the system rose at accelerated rate at high temperatures above $\sim 1373$ K as demonstrated in Fig. 1(b). This observation proves that temperature plays an important role in the transformation process of HDPE as it provides the kinetic energy to break the bonds C--H bonds which is prerequisite \ch{H2} formation at temperatures above 1396 K This point is even more signified when we consider that the C--C bond formation energy is 348 kJ$/$mol while C--H bonds have a considerably higher formation energy of 413kJ/mol. The role of such high temperature in \ch{H2} generation seems to be critical. For instance, a detailed analytical model of HDPE pyrolysis and the corresponding experiments did not predict any \ch{H2} generation at temperatures below $\sim 973$ K [7,25]. \par
Since the rise in temperature was coupled with a sharp decrease in the density, we expect that depropagation reactions further degrade the HDPE chain to form lighter molecules. The depropagation reactions in which small alkenes are generated from the decomposition of larger alkane radicals are well known to be dominant in dilute polymeric systems [26]. In the case of HDPE, the depropagation reaction generates new shorter monomers from the freshly formed radicals. During the depropagation reaction, the termination occurs when a pair of radicals interact to reform a new molecule. This process is illustrated by the set of Equations 1--5. Equation 1 shows free radical formation by breakdown while Equations 2 and 3 show free radical transfer and finally Equation 4 shows the free radical termination by disproportionation while Equation 5 shows the radical termination by combination.\par

\begin{equation}
\ch{R} \textendash \ch{R} \rightarrow 2 \ch{R^.},
\end{equation}

\begin{equation}
2\ch{R^. } \rightarrow \ch{R} \textendash \ch{C^.H-CH2-CH2} \textendash \ch{R} \rightarrow \ch{R^. }+\ch{CH2=CH-CH2-R},
\end{equation}

\begin{equation}
\ch{R^. }+\ch{R-CH2-R} \rightarrow \ch{R-C^.H-RH} \rightarrow \ch{R-C^.H=CH2-R^.},
\end{equation}

\begin{equation}
\ch{R^. }+\ch{R^. } \rightarrow \ch{R-CH=CH2} + \ch{RH} ,
\end{equation}

\begin{equation}
\ch{R^. }+\ch{R^. } \rightarrow \ch{R} \textendash \ch{R}.
\end{equation}

As in Table 1, we see that generation of hydrogen molecules was initiated at the temperature of 1396 K which was preceded by the generation of \ch{CH_n} molecules at 988 K. Furthermore the number of hydrogen and \ch{CHn} molecules increased dramatically when temperature passed above $\sim 1373$ K. This trend is in agreement with earlier experiments where high level of hydrogen release from HDPE at high temperatures was reported [27]. In this stage, the hydrogen generation stemmed from the thermal cracking reactions of ethane and ethylene described in the Equations (6--9):\

Initiation\
\begin{equation}
\ch{C2H6} \rightarrow 2\ch{CH3},
\end{equation}

Propagation\
\begin{equation}
  \begin{cases}
\ch{CH3} + \ch{C2H6} \rightarrow \ch{CH4} + \ch{C2H5},\\
\ch{C2H5} \rightarrow \ch{C2H4} + \ch{H},\\
\ch{H} + \ch{C2H6} \rightarrow \ch{H2} + \ch{C2H5},
  \end{cases}
\end{equation}

Termination\
\begin{equation}
\ch{C2H5} + \ch{C2H5} \rightarrow \ch{C2H6} + \ch{C2H4},
\end{equation}

\begin{equation}
\ch{C2H6} + 1600 \text{kcal/kmol} \rightarrow \ch{C2H2} + \ch{H2}.
\end{equation}

To cross examine the \ch{H2} and \ch{CH_n} generation at higher temperature, we simulated the HDPE degradation process at 1823 K using ab initio molecular dynamics as implemented in SIESTA code [28]. Ab initio simulation does not depend on predefined force-field and thus its accuracy is completely temperature independent. Fig. 2(a) schematically presents the initial structure of the HDPE chain molecule. After the system was equilibrated, we specified the output molecules by probing the charge densities surrounding every individual atom to establish bond connectivity as demonstrated in Fig. 2(b). We found that total number of 36 light molecules of \ch{H2} and \ch{CH_n} were generated by the end of the run which is in general agreement with the classical molecular dynamics simulation. It appears that the generation of \ch{H2} and \ch{CH_n} is facilitated by the formation of double and triple C--C bonds in \ch{C2H_n} and \ch{C3H_n} at higher when temperature is sufficiently high.

       \section{CONCLUSIONS}\FloatBarrier
In conclusion, we conducted molecular dynamics simulations to understand thermal transformation of HDPE at high temperatures. Our results suggest that at high temperature of steel making, thermal degradation of HDPE at 1823 K results in evolution of important gaseous reductants (\ch{H2}, \ch{CH_n}). With increasing the temperature from 1396 K to 1823 K, the number of hydrogen and \ch{CH_n} molecules increased dramatically. These results suggest HDPE as a good reductant candidate to generate significant reducing gases including hydrogen in steelmaking applications.
             
       \section{CONFLICTS OF INTEREST}
The authors declare that there is no conflict of interest.
       
      \section{ACKNOWLEDGMENTS}
This work was supported by Australian Research Council through Grant No. FT0992021. The computational facility was provided by Intersect Australia Ltd.

       \section{References}
[1]	H.D. Lu, Y. Hu, J.F. Xiao, Z.Z. Wang, Z.Y. Chen and W.C. Fan, Magnesium Hydroxide Sulfate Hydrate Whisker Flame Retardant Polyethylene/Montmorillonite Nanocomposites, J. Mater. Sci. 41 (2006) 363--367.\par
[2]	K. Kaygusuz and A. Sari, High Density Polyethylene/Paraffin Composites as Form-Stable Phase Change Material for Thermal Energy Storage, Energy Source. A 29 (2007) 261--270.\par
[3]	K.S. Shah, R.C. Jain, V. Shrinet, A.K. Singh and D.P. Bharambe, High Density Polyethylene (HDPE) Clay Nanocomposite for Dielectric Applications, IEEE T. Dielect. El. In. 16 (2009) 853--861.\par
[4]	V. Sahajwalla, M. Rahman, R. Khanna, N. Saha-Chaudhury, P. O'Kane, C. Skidmore and D. Knights, Recycling Waste Plastics in EAF Steelmaking: Carbon/Slag Interactions of HDPE-Coke Blends, Steel Res. Int. 80 (2009) 535--543.\par
[5]	V. Sahajwalla, M. Zaharia, S. Kongkarat, R. Khanna, M. Rahman, N. Saha-Chaudhury, P. O'Kane, J. Dicker, C. Skidmore and D. Knights, Recycling End-of-Life Polymers in an Electric Arc Furnace Steelmaking Process: Fundamentals of Polymer Reactions with Slag and Metal, Energy Fuel. 26 (2012) 58--66.\par
[6]	M.D. Wallis and S.K. Bhatia, Thermal Degradation of High Density Polyethylene in a Reactive Extruder, Polym. Degrad. Stabil. 92 (2007) 1721--1729.\par
[7]	S. Ali, A.A. Garforth, D.H. Harris, D.J. Rawlence and Y. Uemichi, Polymer Waste Recycling Over "Used" Catalysts, Catal. Today 75 (2002) 247--255.\par
[8]	R. Bagri and P.T. Williams, Catalytic Pyrolysis of Polyethylene, J. Anal.  Appl. Pyrol. 63 (2002) 29--41.\par
[9]	P. Budrugeac, Theory and Practice in the Thermoanalytical Kinetics of Complex Processes: Application for the Isothermal and Non-Isothermal Thermal Degradation of HDPE, Thermochim. Acta 500 (2010) 30--37.\par
[10]	A. Brems, Dewil, R. , Baeyens, J. and Zhang, R, Gasification of Plastic Waste as Waste-to-Energy or Waste-to-Syngas Recovery Route., Nat. Sci. 5 (2013) 695--704.\par
[11]	M.P. Pandey and C.S. Kim, Lignin Depolymerization and Conversion: A Review of Thermochemical Methods, Chem. Eng. Technol. 34 (2011) 29--41.\par
[12]	E. Ranzi, M. Dente, T. Faravelli, G. Bozzano, S. Fabini, R. Nava, V. Cozzani and L. Tognotti, Kinetic Modeling of Polyethylene and Polypropylene Thermal Degradation, J. Anal. Appl. Pyrol. 40 (1997) 305--319.\par
[13]	P.E. Sanchez-Jimenez, L.A. Perez-Maqueda, A. Perejon and J.M. Criado, A New Model for the Kinetic Analysis of Thermal Degradation of Polymers Driven by Random Scission, Polym. Degrad. Stabil. 95 (2010) 733--739.\par
[14]	E. Ranzi, M. Dente, A. Goldaniga, G. Bozzano and T. Faravelli, Lumping Procedures in Detailed Kinetic Modeling of Gasification, Pyrolysis, Partial Oxidation and Combustion of Hydrocarbon Mixtures, Prog. Energy Combust. Sci. 27 (2001) 99--139.\par
[15]	S. Plimpton, Fast Parallel Algorithms for Short-Range Molecular-dynamics, J. Comput. Phys. 117 (1995) 1--19.\par
[16]	A.C.T. van Duin, S. Dasgupta, F. Lorant and W.A. Goddard, ReaxFF: A Reactive Force Field for Hydrocarbons, J. Phys. Chem. A 105 (2001) 9396--9409.\par
[17]	A. Strachan, E.M. Kober, A.C.T. van Duin, J. Oxgaard and W.A. Goddard, Thermal Decomposition of RDX from Reactive Molecular Dynamics, J. Chem. Phys. 122 (2005) 054502.\par
[18]	P. Hunenberger, in Advanced Computer Simulation Approaches for Soft Matter Sciences I, edited by C. Holm and K. Kremer (2005), Vol. 173, pp. 105--147.\par
[19]	T. Schneider and E. Stoll, Molecular-Dynamics Study Of A 3-Dimensional One-Component Model for Distortive Phase-Transitions, Phys. Rev. B 17 (1978) 1302--1322.\par
[20]	H.J.C. Berendsen, J.P.M. Postma, W.F. Vangunsteren, A. Dinola and J.R. Haak, Molecular-dynamics with Coupling to an External Bath, J. Chem. Phys. 81 (1984) 3684--3690.\par
[21]	 Typical Engineering Properties of High Density Polyethylene, (2014) \url{http://www.ineos.com/Global/Olefins\%20and\%20Polymers\%20USA/Products/Technical\%20information/INEOS\%20Typical\%20Engineering\%20Properties\%20of\%20HDPE.pdf}, Accessed on August 20 2014.\par
[22]	S. Moss and H. Zweifel, Degradation and Stabilization of High-density Polyethylene During Multiple Extrusions, Polym. Degrad. Stabil. 25 (1989) 217--245.\par
[23]	J. Michal, J. Mitera and S. Tardon, Toxicity of Thermal Degradation Products of Polyethylene and Polypropylene, Fire Mater. 1 (1976) 160--168.\par
[24]	J.A. Conesa, A. Marcilla, R. Font and J.A. Caballero, Thermogravimetric Studies on the Thermal Decomposition of Polyethylene, J. Anal. Appl. Pyrol. 36 (1996) 1--15.\par
[25]	J.F. Mastral, C. Berrueco and J. Ceamanos, Modelling of the Pyrolysis of High Density Polyethylene - Product Distribution in a Fluidized Bed Reactor, J. Anal. Appl. Pyrol. 79 (2007) 313--322.\par
[26]	W.G. Barb, Role of Depropagation Reactions in the Copolymerization of Styrene and Sulfur Dioxide, J. Polym. Sci. 10 (1953) 49--62.\par
[27]	J.R. Dankwah, P. Koshy, N.M. Saha-Chaudhury, P. O'Kane, C. Skidmore, D. Knights and V. Sahajwalla, Reduction of FeO in EAF Steelmaking Slag by Metallurgical Coke and Waste Plastics Blends, Isij Int. 51 (2011) 498--507.\par
[28]	J.M. Soler, E. Artacho, J.D. Gale, A. Garcia, J. Junquera, P. Ordejon and D. Sanchez-Portal, The SIESTA Method for Ab Initio Order-N Materials Simulation, J. Phys.: Condens. Matter 14 (2002) 2745--2779.\par

\end{document}